\documentclass[12pt]{article}

\usepackage{scicite}

\usepackage{times}

\usepackage{graphicx}
\usepackage{textcomp}

\newenvironment{sciabstract}{%
\begin{quote} \bf}
{\end{quote}}

\title{Monolithic shape-programmable dielectric liquid crystal elastomer actuators} 

\author
{Zoey S. Davidson$^{1\ast}$, Hamed Shahsavan$^{1}$, Amirreza Aghakhani$^{1}$, Yubing Guo$^{1}$,\\
Lindsey Hines$^{1}$,Yu Xia$^{2}$, Shu Yang$^{2}$, Metin Sitti$^{1,3}$\\
\\
\normalsize{$^{1}$Department of Physical Intelligence, Max Planck Institute for Intelligent Systems,}\\
\normalsize{Stuttgart, Germany}\\
\normalsize{$^{2}$Department of Materials Science and Engineering, University of Pennsylvania,}\\
\normalsize{Philadelphia, USA}\\
\normalsize{$^{3}$School of Medicine and School of Engineering, Ko\c{c} University,}\\
\normalsize{Istanbul, Turkey}\\
\\
\normalsize{$^\ast$To whom correspondence should be addressed; E-mail:  davidson@is.mpg.de.}
}

\date{}

\begin{document}

\baselineskip24pt

\maketitle 

\begin{sciabstract}

Macroscale robotic systems have demonstrated great capabilities of high speed, precise, and agile functions. However, the ability of soft robots to perform complex tasks, especially in centimeter and millimeter scale, remains limited due to the unavailability of fast, energy-efficient soft actuators that can programmably change shape. Here, we combine desirable characteristics from two distinct active materials: fast and efficient actuation from dielectric elastomers and facile shape programmability from liquid crystal elastomers into a single shape changing electrical actuator. Uniaxially aligned monoliths achieve strain rates over 120\%/s with energy conversion efficiency of 20\% while moving loads over 700 times the actuator weight. The combined actuator technology offers unprecedented opportunities towards miniaturization with precision, efficiency, and more degrees of freedom for applications in soft robotics and beyond.

\end{sciabstract}

Force generation, efficiency, strength-to-weight ratio, work capacity, and shape programmability will be key for the next generation of soft robots to perform complex functions. Despite significant advances in robots, including gymnastic feats \cite{PopeStickmanHumanScale2018}, the underlying rigid actuation mechanisms, use of electric motors or hydraulic and pneumatic actuators, remain relatively unchanged and potentially hinder their miniaturization and, more importantly, their use in human collaborative environments \cite{MaddenMobileRobotsMotor2007}.  Efficient and programmable soft actuators, like an artificial muscle, would significantly increase the capabilities and potential applications of soft robotic systems in aerospace, industrial, or medical technologies \cite{MirvakiliArtificialMusclesMechanisms2018, WhitesidesSoftRobotics2018,MajidiSoftMatterEngineering2018}. Among many soft actuation mechanisms that have been explored, dielectric elastomer (DE) actuators appear promising and even outperform skeletal muscle in some aspects \cite{Gusurveydielectricelastomer2017,BrochuAdvancesDielectricElastomers2010,DudutaRealizingpotentialdielectric2019,AcomeHydraulicallyamplifiedselfhealing2018}. Separately, liquid crystal elastomers (LCEs) have demonstrated reversible large mechanical deformation by thermal and optical actuation. Recent advances in photoalignment and top-down microfabrication techniques have enabled pre-programming of LC alignment in microdomains for complex shape morphing \cite{deGennesmuscleartificielsemirapide1997,WareVoxelatedliquidcrystal2015a,WhiteProgrammableadaptivemechanics2015a}. However, both actuator types have their drawbacks: DE films need to be pre-stretched, making it difficult to program local actuation behaviors microscopically  \cite{HajiesmailiReconfigurableshapemorphingdielectric2019}.  Meanwhile, directly converting electrical energy to mechanical work utilizing LCEs has, until now, remained limited due to the small strain generated \cite{LehmannGiantlateralelectrostriction2001a,CNematicAnisotropicLiquidCrystal2003,GuinElectricalControlShape2018,VerduzcoShapeshiftingliquidcrystals2015,PetschMuscularMEMSengineering2016,SpillmannElectricallyInducedTwist2016}.

Typically, DE actuators function by electrostatic attraction between two compliant electrodes coated on opposing sides of an isotropic DE to form a variable resistor-capacitor \cite{PelrineHighSpeedElectricallyActuated2000}. High voltage applied to the compliant electrodes induces an electrostatic pressure called Maxwell stress. The electrical actuation mechanism can result in much higher operating efficiency (ratio of mechanical work to input electrical energy) and faster actuation speed than those of LCEs \cite{DudutaRealizingpotentialdielectric2019,AcomeHydraulicallyamplifiedselfhealing2018}. Besides functioning as soft linear actuators, DE actuators could be applied to grippers, haptic, or optical devices, which however require complex shape change \cite{BrochuAdvancesDielectricElastomers2010}. Despite some impressive demonstrations, DE actuators have not yet gained widespread use in soft robotics due in part to their need for pre-strech and their difficulty in fabricating devices with complex deformation profiles \cite{HajiesmailiReconfigurableshapemorphingdielectric2019}. To overcome these challenges and expand the applications of DE actuators, it requires material innovation to create high performance DEs with programmability \cite{BiggsJamesElectroactivePolymersDevelopments2013, OprisDorinaM.PolarElastomersNovel2017}.

LCEs are rubbery polymers with anisotropic bulk properties imparted by their constituent molecular anisotropy. Most prior work on LCE actuation has focused on thermal or light driven mechanisms, which temporarily disrupt the anisotropic molecular order, known as the director field ($\mathbf{n}$), thereby creating internal stresses and anisotropic bulk deformation \cite{WareVoxelatedliquidcrystal2015a, GelebartMakingwavesphotoactive2017a, PetschMuscularMEMSengineering2016}. The local LCE director field can be programmed to create complex shape change when actuated. However, light actuation is inefficient and thermal actuation is both slow and inefficient thus they are poorly suited to integration with robotics applications. Indeed, direct electrical actuation of LCEs is highly sought after \cite{VerduzcoShapeshiftingliquidcrystals2015}. A few previous studies have demonstrated electrical LCE actuation by coupling an electric field to the molecular dielectric anisotropy or sometimes intrinsic polarization of the LCE or LCE composite; an electric field drives molecular reorientation to create bulk strains. However, these methods require elevated temperatures, carbon nanotube dopants, or produce small actuation strains \cite{LehmannGiantlateralelectrostriction2001a, GuinElectricalControlShape2018, FengWeiOscillatingChiralNematic2018, CNematicAnisotropicLiquidCrystal2003, OkamotoLargeelectromechanicaleffect2011}. LCEs' elastic anisotropy and thus Poisson's ratio anisotropy makes them an ideal material for shape programmable DEs. In this report, we study how LCEs perform as a DE actuator, which we refer to as dielectric liquid crystal elastomer actuators (DLCEA, Fig. 1B). By pre-programming LC alignment in local domains, we achieve electric driven actuation and shape morphing at room temperature, and demonstrate large, fast, and forceful strains.

The LCE films are fabricated in a two-step process recently developed by some of the authors \cite{AharoniUniversalinversedesign2018, XiaInstantLockingMolecular2018, Seesupplementarymaterials}. Briefly, an oligomer is synthesized prior to LCE film fabrication by a thiol-acrylate click reaction; a common diacrylate reactive liquid crystal monomer is chain extended by Michael addition with a dithiol linker molecule. The exact component ratios, choice of monomer, and dithiol linker can all be tuned to adjust the specific mechanical properties of the final LCE film \cite{XiaInstantLockingMolecular2018, GodmanSynthesisElastomericLiquid2017}. We produce large areas of well ordered uniaxial LCE (fig. S1,S2) with giant elastic anisotropy (Fig. 1C). Throughout this work, we actuate DLCEAs at room temperature and only in the linear regime where strains do not induce LCE director reorientation (fig. S3). We can also locally program the LCE director field by photoalignment to create a spatially programmed command surface to locally orient the LCE director \cite{YaroshchukPhotoalignmentliquidcrystals2011}. Finally, we apply compliant grease electrodes to both sides of the LCE film to create DLCEA devices (fig. S4).

To characterize the fundamental properties of DLCEAs, we first made uniaxially aligned LCE films. The electrodes coated on uniaxial DLCEAs enables simultaneous measurement of capacitance and stress versus strain applied to the LCE film (Fig.\ 1C). The LCE is more than an order of magnitude stiffer when strains ({\bf u}) are applied parallel to the director, {\bf n}$||${\bf u} (16~MPa), than when strains are applied perpendicular to the director, {\bf n}$\perp${\bf u} (1.2~MPa), which indicates a high degree of elastic anisotropy. Similarly, the difference in slope of the normalized capacitance between the DLCEA devices with different director orientation indicates anisotropy in the Poisson's ratio. The DLCEA capacitance is proportional to the area of the film coated by the electrode and inversely proportional to the film thickness. Thus Poisson's ratio anisotropy causes the film's thickness and area to change at different rates depending on the direction of strain relative to the LCE director field (fig. S5)\cite{Seesupplementarymaterials}. Other works with similar LCE chemistry have observed 1D translational crystallinity (smectic or cybotactic) order which may explain the particularly large elastic anisotropy observed in this work \cite{SaedHighstrainactuation2017,GodmanSynthesisElastomericLiquid2017}. When using the DLCEA as a linear actuator, we expect larger Poisson's ratio anisotropy to lead to higher efficiency and lower electric fields required for actuation than compared to a similar isotropic material \cite{HuangLargeunidirectionalactuation2012}. However, reducing viscous loss, indicated here by the hysteresis loop in Fig.\ 1C, is also important for fast and efficient soft actuators.

We then characterized uniaxial DLCEAs in isometric (constant strain, Fig.\ 2A) and isotonic (constant force, Fig.\ 2B) configurations. In isometric tests, we applied initial strains to DLCEA devices and allowed a relaxation period prior to application of high voltages (fig.\ S6A). From the active stress, we observe two relationships between the strain, voltage applied, and active stresses, which fit with the model of Maxwell stress, $p_{\mathrm{es}}\propto V^2/d^2$, where  $V$ is the applied voltage, and $d$ is the LCE film thickness. First, actuation at larger initial strains produces higher active nominal stresses; the isometric prestrain results in thinning of the LCE and thus higher Maxwell stress for a given voltage (Fig.\ 2A). This is an advantage of LCEs compared to the behavior of an isotropic DE where typically strain stiffening offsets the increasing Maxwell stress. With an LCE, it appears that additional strain thins the material at a rate fast enough to offset the stiffening. We also observe that, for each fixed strain, the active nominal stress during isometric tests increases quadratically with increasing voltage (fig.\ S6B). At the highest voltages tested, we measured peak active nominal stresses in excess of 50~kPa. However, for the device with the director ${\bf n}||{\bf u}$, the active stresses are relatively much smaller due to the much higher modulus. When held with an isometric strain, the DLCEA behaves like a variable stiffness spring, and in the case of 5\% ${\bf n}\perp{\bf u}$ initial strain and 2~kV actuation voltage, the Maxwell stress induced expansion of the LCE nearly compensates the entirety of the isometric strain induced stress. We also performed isopotential tests in which the DLCEA is strained under a constant voltage. These tests indicate the expected actuation stroke of a loaded DLCEA when a voltage is applied (fig.\ S7).

Next, we take the same DLCEA and perform isotonic tests by suspending varying weights from the free end of the DLCEA to generate constant load forces and initial nominal strains, ${\bf u}$. DLCEAs in the ${\bf n} || {\bf u}$ configuration do not show an appreciable active strain at any initial loading for even the highest voltages tested due to their significantly higher elastic modulus (movie S1). However, DLCEAs with ${\bf n} \perp {\bf u}$ configuration exhibit fast active strains up to 5\% with the application of 3~kV for the heaviest loading tested, 0.27~N, approximately 790 times the bare LCE film weight of 35~mg. We perform isotonic contractile tests by abruptly discharging weighted DLCEA devices and capture the subsequent motion with high speed video (figs. S8A and movie S2). With increasing load and initial strain, the DLCEA capacitance increases, but in all cases, the electrical discharge time, approximately 1~ms, is much faster than the DLCEA elastically contracts, within 60~ms (Fig.\ 2B), which indicates the system is currently limited by the viscoelastic properties of the LCE. We also observed significant viscous losses in the contractions which are apparent from the continued creeping contraction after the initial elastic response (fig. S8B). From the DLCEA contraction trajectory, we compute fundamental performance metrics both for the pure elastic response and for the extended creeping contraction (Fig. 2C, fig. S8C). Compared to other LCE actuators, our reporting of actuation efficiency of approximately 20\% is remarkable; actuation efficiency in LCEs has not, to our knowledge, been reported before for LCEs performing external work. We expect LCE actuation studies using light or thermal methods to achieve energy conversion efficiency of 1\% or less.

To achieve complex shape actuation, LCEs typically function by programming spatially varying in-plane contractile strain. However, the DE actuator mechanism we utilize here generates in-plane expansion strains. Thus boundary conditions play a significant role in determining the realized DLCEA shape change. To better understand the role of boundary conditions on DLCEAs, we performed a fundamental characterization of the buckling effect caused by the elastomer's expansion between fixed boundaries (Fig. 3A,B, movie S3). The buckling amplitude increases with increasing voltage and, at 2.5~kV, creates out-of-plane peak-to-peak strokes greater than 1800\% of the LCE film thickness of approximately 80~\textmu m (Fig. 3C). Actuation speed is another important characteristic for potential DLCEA applications such as haptic interfaces. We applied a sinusoidally varying 1~kV potential to measure the change in actuation amplitude as a function of the applied frequency (Fig. 3D, fig S9). The actuation amplitude decays exponentially with the frequency, but would still be a perceivable 50~\textmu m at 30~Hz and 1~kV.

We designed spatially varying LCE director configurations, aiming to demonstrate ability to pre-program complex patterns in two dimensions (2D), followed by electrical actuation of the films into three dimensional (3D) forms \cite{WareVoxelatedliquidcrystal2015a, AharoniUniversalinversedesign2018}. Depending on the programmed director field, the LCE film buckles out-of-plane with locally positive (Fig.\ 4A) or negative Gaussian curvature (Fig.\ 4B). We created a pixelated array of topological defects by spatially programming light polarization with a pattern of linear film polarizers (Fig.\ 4C) to locally orient the LCE director as in Fig. 4D. The director forms a lattice of radial and azimuthal type defects that, when electrically actuated, buckle out-of-plane due to incompatible in-plane strains (Fig 4E, movie S4). We measured the height of the DLCEA surface in the discharged (0~V) and 2.5~kV actuated (Fig.\ 4F) states. To complete these height field measurements, the device stably held its actuated shape at 2.5~kV for over 3 h while drawing less than 1~\textmu A thus demonstrating stability and low power consumption to maintain the shape change. The locally programmed height change and accompanying change in Gaussian curvature is clearly visible from circular traces enclosing the central radial type defect (Fig.\ 4G). The out-of-plane buckling creates peak-to-peak height differences of over 1600~\textmu m -- a 2000\% growth from the initial film thickness of approximately 80~\textmu m. The lower right corner of the actuated DLCEA in Fig. 4F shows it is possible for the defects to buckle both up and down. Though, in principle, the buckled shape of the DLCEA is multistable, we only ever observe a single actuated state for each sample \cite{WareVoxelatedliquidcrystal2015a}. The symmetry that would enable multistability is likely broken by gravity during testing or by directional exposure to heat and UV light during device fabrication. This demonstration of local change in the Gaussian curvature indicates our method can be generalized to realize a large variety of flat to curved surface programmed shape changes. In addition to programming varying in-plane director orientations, it is also possible for the LCE director orientation to vary through the film thickness. As a demonstration of through-thickness LCE programming, we have also assembled DLCEAs with a twisted nematic LCE configuration where the director rotates by 90 degrees from one surface of the film to the other. When electrically actuated, this twisted nematic DLCEA produces large electrically activated twisting motions (fig. S10) \cite{Seesupplementarymaterials}.

Here, by combining desirable characteristics of DEs and LCEs in a single material platform, we demonstrate superior actuation performance from electrically actuated DLCEAs, including high efficiency (20\%), fast actuation speed (120\%/s) and programmable shape change from 2D to 3D with more than 1800\% out-of-plane stroke. The insights on integration of active materials with top-down microfabrication techniques and electroactuation mechanism presented here will offer exciting opportunities when coupling DLCEAs with 3D printing, origami/kirigami actuation strategies, as well as distributed control systems toward creating multi-functional soft robots in a scalable fashion at a low framing cost. The electroactuation mechanism can also be applied to other technologies, including energy harvesting and storage, medical devices, wearable technology, and aerospace. Furthermore, fast and dynamic modulation could be useful in displays and optical applications.

\section*{Acknowledgments}
The authors thank Matthew Millard, Steve W. Heim, and members of the Physical Intelligence Department for helpful discussions. Z.S.D. and Y.G are supported by the Alexander von Humboldt Foundation. H.S. is supported by Natural Sciences and Engineering Research Council of Canada. S.Y. also wishes to acknowledge partial support from National Science Foundation (NSF)/EFRI-ODISSEI grant, \#EFRI-1331583. A.A., L.H., and M.S. are supported by the Max Planck Society.

\section*{Supplementary materials}
Materials and Methods\\
Supplementary Text\\
Figs. S1 to S10\\
Movies S1 to S4\\

\clearpage

\includegraphics[width=150mm]{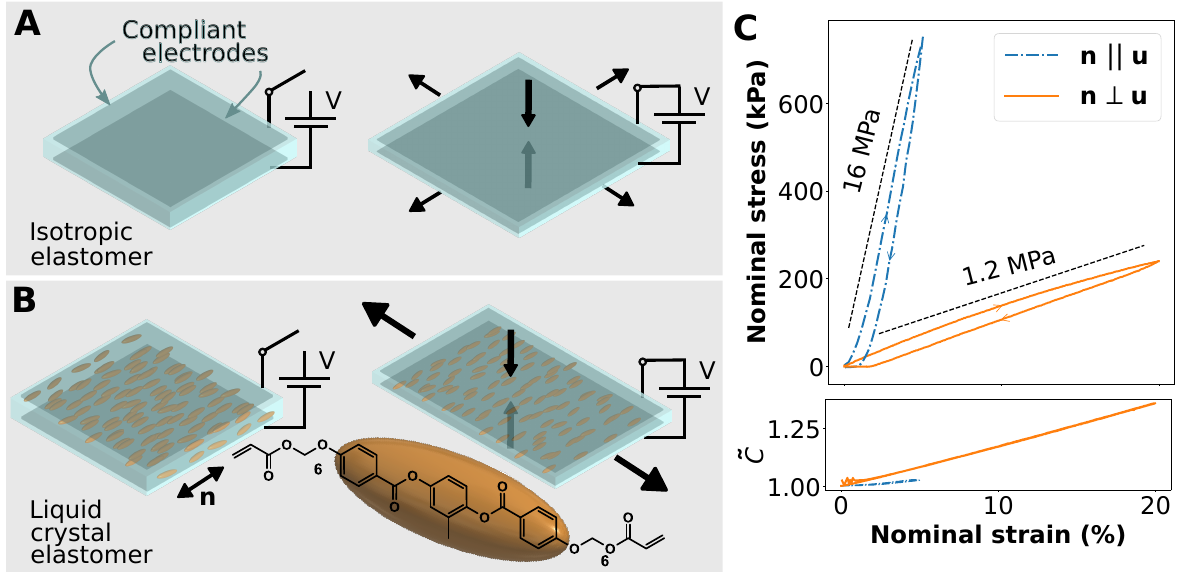}

\noindent {\bf Fig. 1. Device schematic, mechanical, and electrical characterization} 

\noindent  (\textbf{A}) Schematic of a traditional isotropic DE actuator in off and on states. (\textbf{B}) Schematic of a uniaxial aligned DLCEA in off and on states. Liquid crystal molecular alignment, the director {\bf n}, is indicated by double headed arrow and defines the stiffer direction of the LCE. When actuated by a voltage, $V$, the material thins and stretches perpendicular to the alignment greater than parallel to the director. (\textbf{C}) The DLCEA mechanical stress and normalized  capacitive ($\widetilde{C}$) response to strain over the DLCEA linear regime are characterized at a strain rate $1\times10^{-3}$/s.

\clearpage

\includegraphics[width=150mm]{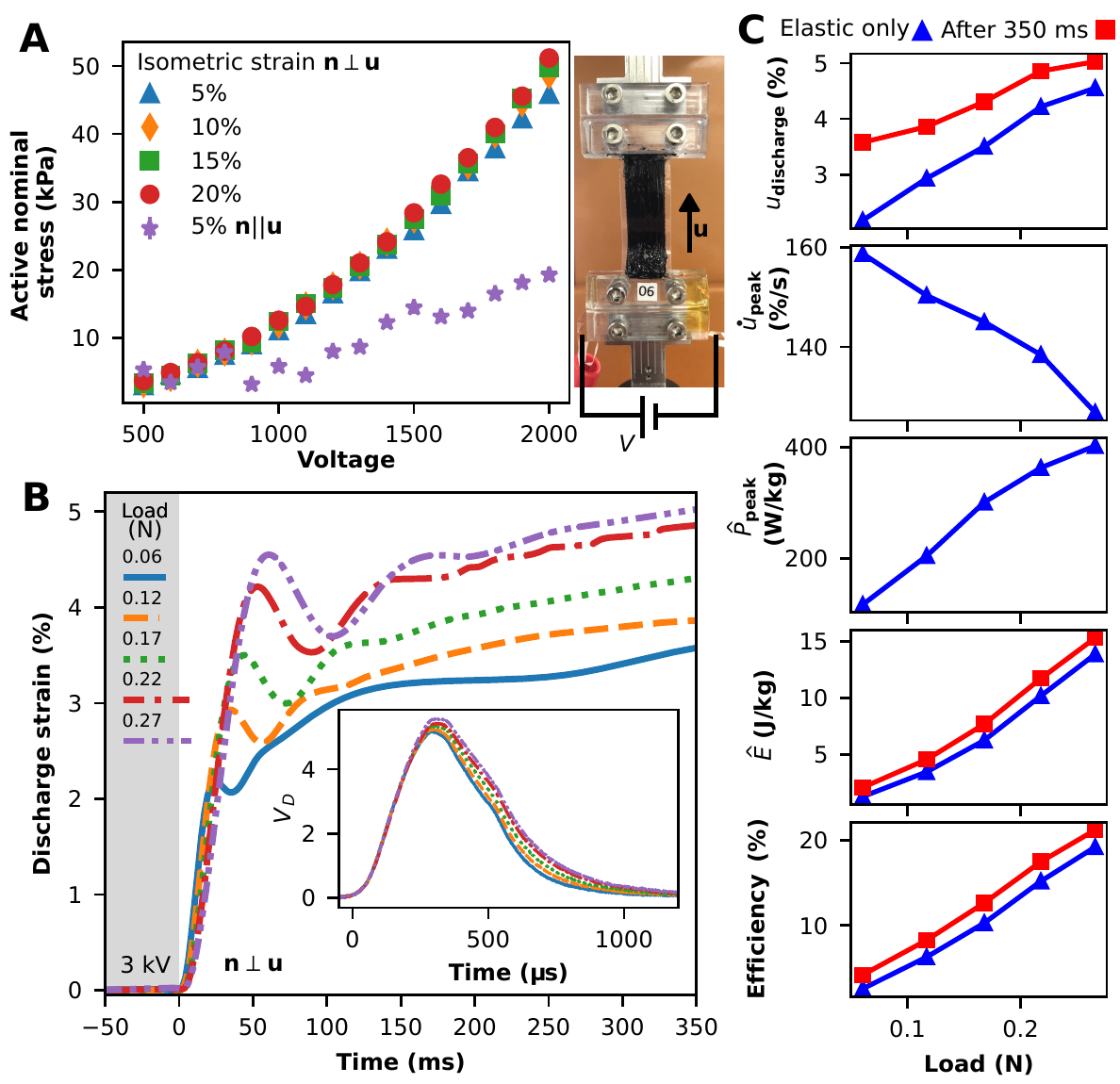}

\noindent {\bf Fig. 2. Characterization of uniaxial nematic DLCEA}

\noindent (\textbf{A}) Isometric (constant strain) and (\textbf{B,C}) isotonic (constant force) tests demonstrate the capabilities of a DLCEA actuator device. (\textbf{A}) Measured active nominal stress with various initial isometric strains (${\bf u}$) for devices assembled with the LCE director {\bf n}$\perp${\bf u} and {\bf n}$||${\bf u}. And a photograph of an assembled DLCEA device with {\bf n}$\perp${\bf u}.  (\textbf{B}) Contractile discharge strain trajectories under various loads measured by high speed camera with actuation voltages of 3~kV. Inset, the corresponding measurements of electrical discharge. (\textbf{C}) Fundamental actuator characteristics are computed from the contraction trajectory and measurement of the discharge current found in (\textbf{B}), including strain ($u$), peak strain rate ($\dot{u}_{\mathrm{peak}}$), peak specific power ($\hat{P}_{\mathrm{peak}}$), specific energy ($\hat{E}$), and efficiency.

\clearpage

\includegraphics[width=150mm]{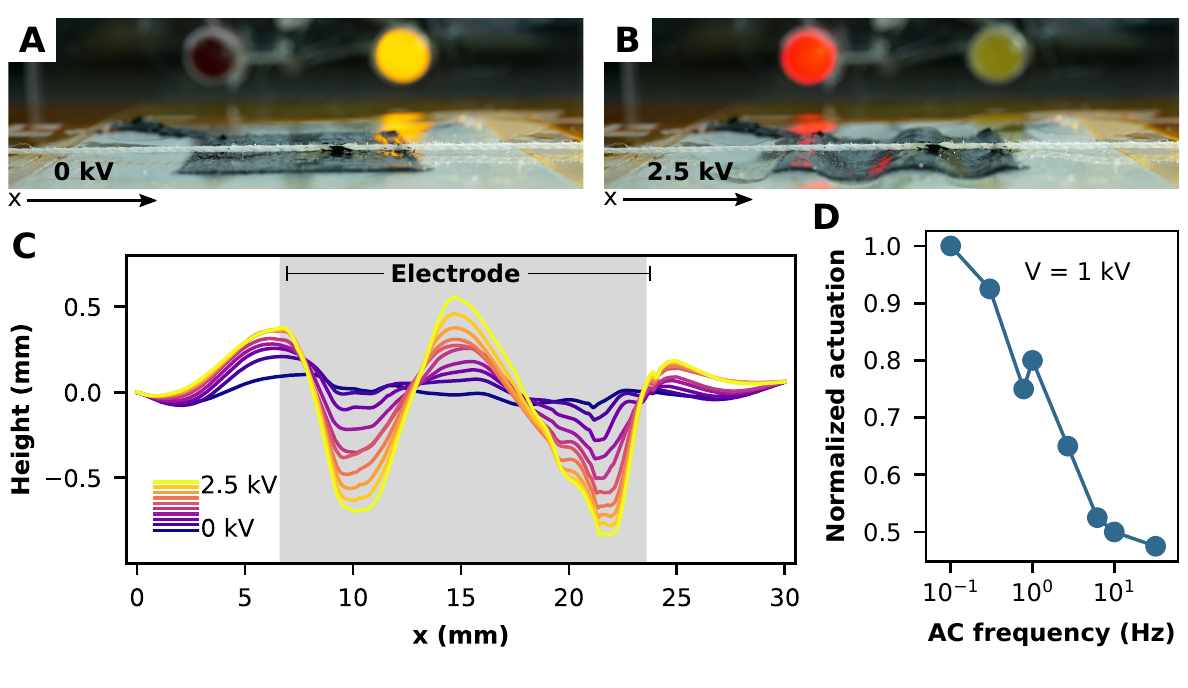}

\noindent{\bf Fig. 3. Uniaxial out-of-plane buckling DLCEA } 

\noindent (\textbf{A}) Off and (\textbf{B}) on states of a uniaxial DLCEA device with fixed boundary condition. Expansion along the soft direction creates out-of-plane buckling, which displaces a fine thread held taught across the surface. (\textbf{C}) Experimental measurement of buckling as a function of applied voltage. (\textbf{D}) Frequency response of buckling uniaxial DLCEA at 1 kV. The  0.1~Hz actuation amplitude is approximately 130 \textmu m.

\clearpage

\includegraphics[width=150mm]{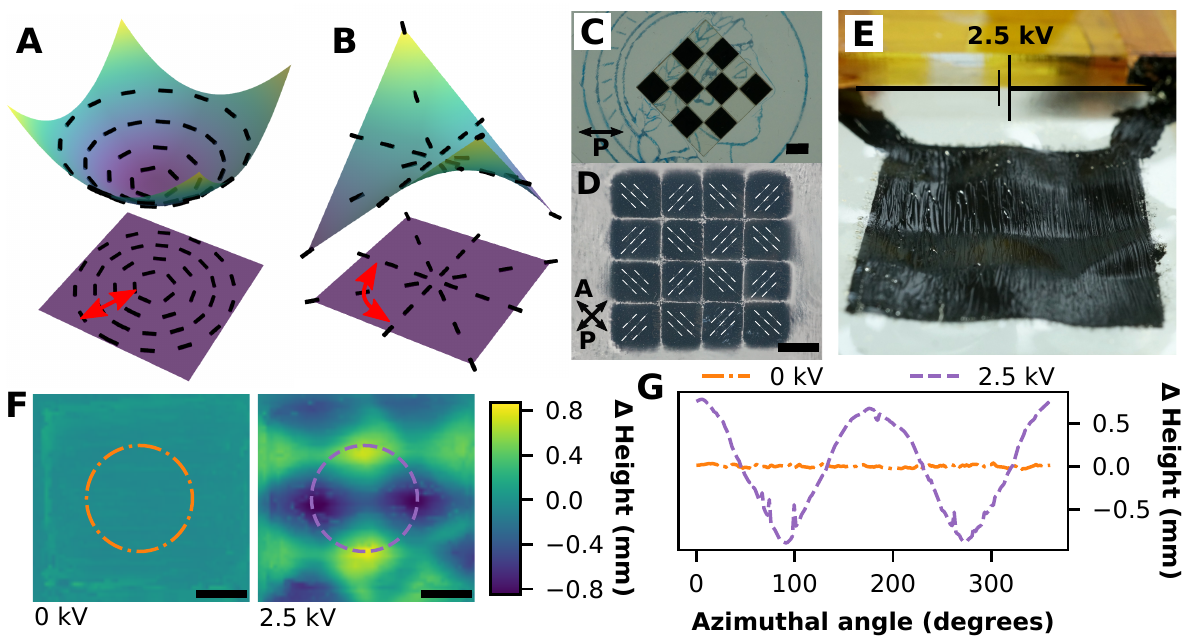}

\noindent {\bf Fig. 4. Pixelated dielectric liquid crystal elastomer actuator} 

\noindent Programmed shape actuation, such as a dimple pattern deformation, is possible by patterning the director configuration into an azimuthal-radial defect lattice. (\textbf{A}) Azimuthal defect types deform into positive Gaussian curvature and (\textbf{B}) radial defect types deform into negative (saddle-like) Gaussian curvature.  (\textbf{C}) The defects are patterned using a pixelated array of polarizing film with the designed local orientations.  (\textbf{D}) Viewed through crossed polarizers, the fabricated LCE film has pixelated uniaxial alignment, indicated by dashed white lines, forming a defect lattice surrounded by unaligned LCE. (\textbf{E}) When charged to 2.5~kV, there is a large visible deformation of the surface. (\textbf{F}) The height map of the grease covered LCE is nearly flat with no charge and varies over 1.6~mm when charged to 2.5~kV. The dash-dot and dash circles in (\textbf{F}) are traces of height depicted in (\textbf{G}). The change from approximately constant height to a sinusoidally varying height indicates a change in sign of the local Gaussian curvature. All scale bars are 4~mm.

\clearpage

\section*{Materials and Methods}

\subsection*{Materials}

1,5-pentanedithiol (1,5-PDT, $>$ 99\%), 1,8-diazabicycloundec-7-ene (DBU), butylatedhydroxytoluene (BHT), and magnesium sulfate (MgSO$_4$, anhydrous powder) were purchased from Sigma Aldrich and used as received. Hydrochloric acid (HCl), Dimethylformamide (DMF), and Dichloromethane (DCM) were purchased from Fischer Scientific. Photoinitiator, 2,2-dimethoxy-2-phenylacetophenone (DMPA) was purchased from Toronto Research Chemicals. Brilliant Yellow (BY) was purchased from Tokyo Chemical Industry. The LC monomer, 1,4-bis-[4-(6-acryloyloxy-hexyloxy)benzoyloxy]-2-methylbenzene (RM82, $>$ 95\%), was purchased from Wilshire Technologies Inc.\ and used without further purification. Conductive carbon grease, Nyogel 756G, was purchased from Newgate Simms.

\subsection*{LCE oligomer synthesis}

We fabricate LCE films in a two step process recently developed by some of the authors \cite{AharoniUniversalinversedesign2018, XiaInstantLockingMolecular2018}. An oligomer is synthesized prior to LCE film fabrication by a thiol-acrylate click reaction; reactive liquid crystal monomer, RM82, is chain extended by Michael addition with 1,5-pentanedithiol. In a typical synthesis, 12.5~g of RM82 are mixed with 5.06~g of 1,5-PDT in 120~mL of DCM with 3 drops of DBU catalyst. After 16 hours stirring at room temperature, the solution is rinsed in a separation funnel with 1M HCl, 0.1M HCl, and deionized water successively. The DCM and product mixture is then dried with 25~g MgSO$_4$ for 30 minutes, which is then filtered. 50~mg of BHT is added to the clear DCM and product mixture prior to rotary evaporation and direct vacuum until a thick white oligomer remains. The oligomer is stored at $-$30~$^{\circ}$C for up to two months.

\subsection*{Substrate preparation}

Glass slides, typically 5 cm by 5 cm and 8 cm by 10 cm, are cleaned in an ultrasound bath with deionized water, isopropanol, and acetone. Next, the slides are dried with nitrogen and then treated by oxygen plasma. A mixture of 1~wt. \% BY dissolved in DMF is spun coat onto the slides and then dried on a hot plate at 120 $^\circ$C. Spacers cut from polyimide or Mylar plastic, thicknesses 65~\textmu m or 75~\textmu m, are placed along the edges on one of the BY coats sides of the glass slides and then two slides are placed with BY coated faces towards each other. Large paper clips hold the slides together with a polarizer film placed on one side. A custom 447~nm LED light source is used to illuminate the BY coated glass through the polarizer film thereby photo-programming the BY molecule orientations. To program locally varying Gaussian curvature, the polarizer film is cut into pixels then reassembled by hand on a glass slide with the desired orientations (see main text Fig. 4C). The thin layer of BY molecules rearrange perpendicular to the incident light polarization to create a spatially photo-programmed command surface which then locally orients the LCE director \cite{YaroshchukPhotoalignmentliquidcrystals2011}.

\subsection*{LCE fabrication and characterization}

The previously prepared oligomer is melted together with additional RM82 and a small amount of photoinitiator to crosslink the oligomer chains into an elastomer network. In more detail, the LCE oligomer and additional RM82 LC monomer are melted together in 1 to 1 molar ratio assuming that the oligomer consists purely of chains of single unit length RM82 capped by 1,5-PDT on both ends \cite{XiaInstantLockingMolecular2018}. Thus the mixture consists of excess thiol groups, which are likely responsible for a significant portion of the final LCE viscous losses (see main text), but this sparse cross-linking also promotes softness necessary to achieve larger actuation strains. The melt is mixed for only 2 to 3 minutes at 120$~^{\circ}$C and then degassed for approximately 3 minutes in a vacuum oven at 90~$^{\circ}$C. One weight percent DMPA is added and carefully stirred in so as not to reintroduce additional air bubbles.

The isotropic LCE melt is then poured onto the BY coated glass at 80$^\circ$\ C then carefully sandwiched with a second hot BY coated glass substrate. The BY-LCE-BY sandwich is cooled into the nematic phase, approximately 73$^\circ$\ C, and then gradually cooled to room temperature during which time it aligns with the spatial programming imparted by the BY coating and defects arising from the phase transition are annealed. Once the LCE cools to room temperature, it is cured in ultra violet light with an OmniCure S2000 arc source. After exposure to ultra violet light polymerizes the LCE in its programmed state, we immerse the BY-LCE-BY sandwich in water to release the LCE from the BY coated glass substrates.

The final LCE film thickness is confirmed by confocal laser profilometry from regions cut to make actuators (described below). Good alignment and few defects in the LCE are essential characteristics of the film to impart the largest possible elastic anisotropy and achieve optimal materials properties. The high contrast between orientations of the LCE between crossed polarizers is visible in Figure S1. 

The exact component ratios, choice of monomer, and dithiol linker can all be tuned to adjust the specific mechanical properties of the final LCE film. \cite{GodmanSynthesisElastomericLiquid2017, XiaInstantLockingMolecular2018}.

After sheets of LCE are separated from the glass substrates, they are rinsed in water to remove residual BY and dried with nitrogen. The LCE sheets are placed back on glass substrates and carefully inspected to identify defect and bubble free regions for fabrication of DLCEA devices. For uniaxial DLCEA devices, the cleanest identified regions are cut into rectangular pieces typically 14~mm by 34~mm with typical weight of 35~mg. This size film was chosen for ease of handling and for electrical actuation constraints described below. Smaller neighboring regions, 20~mm by 5~mm, are used to initially characterize the stress-strain behavior of the LCE and the large strain behavior.

The edges of the laser cut regions on the larger film are then inspected by laser confocal interferometry (Keyence VK-X200) to confirm the as fabricated height of the LCE films. We find that nominally 65~\textmu m Kapton produces approximately 70~\textmu m LCE films, and that nominally 75~\textmu m Mylar produces approximately 83~\textmu m LCE films. The thicknesses may vary by as much as $\pm$10\% across the as produced LCE sheet (Fig. S2).

\subsection*{DLCEA fabrication}

In the next step of DLCEA fabrication (Fig. S4), we attached compliant electrodes to both sides of the LCE film using an electrically conductive carbon grease, Nyogel 756G, frequently used in other dielectric elastomer systems \cite{RossetFlexiblestretchableelectrodes2013}. To apply the carbon grease, the LCE is first clamped in 3D printed plastic clips with copper tape leads designed to facilitate attaching the device to the equipment used for tests described below and in the main text. Some degree of misalignment in the clamping process is unavoidable. The clipped LCE is held in a laser cut Plexiglas assembly jig and masked with a low adhesive removable tape placed around the edges of the LCE film. The masking adhesive tape creates a border region at the LCE edges with no electrode grease, which serves to prevent shorting of the device during actuation at high voltages. A 2mm gap around the edges was found to be sufficient to prevent shorting at the voltages tested (see Fig. 2a). The grease is applied by painting with a swab applicator and excess is removed with a straight paper edge. The entire Plexiglas jig with LCE film is weighed before and after application of grease electrodes to find the grease weight which is typically 30~mg total for both electrodes of the DLCEA. Other high conductivity electrode materials can achieve better performance while adding much less weight and cross sectional area \cite{DudutaRealizingpotentialdielectric2019}; alternative electrode materials will be studied in future studies of these actuators.

\subsection*{Poisson's ratio anisotropy and DLCEA isometric and isopotential tests}

Throughout this work, we test and actuate DLCEAs at room temperature and only in the linear regime where strains do not induce reorientation of the LCE director. In tests not reported here, we typically find the onset of soft mode deformation (director reorientation) at a critical strain of 45\% to 50\% for ${\bf n} \perp {\bf u}$.

We characterize laser cut uniaxial DLCEAs mechanically and electrically (Fig. 1C, Fig. S2). Tensile tests are performed in a TA Instruments DHR3 and simultaneous capacitance measurements are made with a Hameg 8118 LCR meter. Typically, DLCEAs have a zero strain capacitance of approximately 300~pF. We observe a dependence between the rate of capacitance growth and the direction of tensile strain of the LCE film relative to the director. The capacitance of DLCEAs strained perpendicular to the director grows faster than those strained parallel to the director. Conceptually, we understand this by considering LCEs to be volume conserving; thus, extension in one direction causes contraction in the other directions -- however, contraction is greater perpendicular to the director. In other words, when an LCE film is strained perpendicular to the director, it contracts in thickness, also perpendicular to the director, faster than it contracts in width, which is parallel to the  director. When strained parallel to the director, the LCE contracts equally in both thickness and width (assuming the cross-section is isotropic). We can model how strains affect the DLCEA capacitance.

The capacitance of a parallel plate capacitor (or a DLCEA) is
$
C = \frac{\epsilon_0\epsilon_{\perp}A}{d}
$
where $\epsilon_0$ is the permittivity of free space and $\epsilon_{\perp}$ is the relative permittivity perpendicular to the liquid crystal director (Note: The reactive mesogen used in this work, RM82, is known to have a negative dielectric anisotropy, i.e., $\epsilon_{\perp} > \epsilon_{||}$.). The rectangular area covered by the electrode is $A = S_x\times S_y$, and the film thickness is $d$ (see Fig. S5 for schematic and coordinate system). When the DLCEA is strained by $u_{y}$ (perpendicular to the director), the thickness decreases $u_{z} = -u_{y}\nu_{yz}$ and the width along $x$ decreases $u_{x}=-u_{y}\nu_{yx}$. The thickness and area become $(1+u_{z})d=(1-u_{y}\nu_{yz})d$ and $(1+u_{y})S_y(1+u_{x})S_x=(1+u_{y})(1-u_{y}\nu_{yx})A$, respectively. Thus the capacitance becomes:

$$
C = \frac{\epsilon_0\epsilon_{\perp}(1+u_{y})(1-u_{y}\nu_{yx})A}{(1-u_{y}\nu_{yz})d}.
$$
We next normalize by the capacitance at zero strain and then Taylor expand for small strains, i.e., to only the linear term:
$$
\tilde{C} = \left(1+(1+\nu_{yz}-\nu_{yx})u_{y}\right).
$$
From the system symmetry and inserting into this equation the relationship between the mechanical anisotropy, $E_{y}/\nu_{yx} = E_{x}/\nu_{xy}$, and the assumption that $\nu_{xy} = 0.5$, we obtain values for $\nu_{yx}\approx0.04$ and $\nu_{yz}\approx0.84$. Taken with the measured values of the elastic moduli (Fig. 1C), $E_x$ and $E_y$, the stiffness tensor is fully defined. These values indicate the LCE is surprisingly compressible. However this is unlikely and possibly due to at least two factors: the capacitance Q factor decreases from 32 to 22 at 20\% strain and slight prestrains are unavoidable in measuring the moduli, $E_x$ and $E_y$, of the LCE. Together, these factors lead to an error that may partially account for the apparent compressibility.

Isometric tests were performed by quasistatically increasing applied voltages to pre-strained samples. Following capacitance measurements, the DLCEA still clamped in the rheometer is strained to a fixed amount (5\%, 10\%, 15\%, and 20\%) then allowed to relax for a period until the creep in measured stress is much smaller than the induced stress (Fig. S6A). The actuation voltage (Heinzinger LNC-10kV) is increased 100~V every 15~s starting from 500~V. The middle 5 seconds of each period is sampled to measure the active change in stress due to the applied voltage. The log-active nominal stress versus log-voltage relationship for all isometric strains has a slope of 2.0 (Fig. S6B) following the relationship given by the Maxwell stress equation.

Isopotential tests are performed by straining the DLCEA first with no applied voltage and then 2~kV applied (Fig. S7). The difference in induced stress between the 0~kV and 2~kV curves indicates the expected stroke when the DLCEA is operated as an actuator under a constant load.

\subsection*{DLCEA isotonic tests}

To characterize fundamental properties of the LCE as a muscle-like actuator, we performed tests on DLCEAs strained by a constant gravitational load, ${\bf F_g}$. Weights hung from the DLCEA induce an initial strain which thins the material thus aiding in larger actuation for higher initial loadings. When a voltage is applied to the DLCEA with ${\bf n} \perp {\bf F_g}$, the system adopts a new length due to the changed elastic response of the LCE. The LCE is strain stiffening, so the weight stops when the forces balance; however, after an initial elastic response, the DLCEA continues to creep due to the viscoelastic properties of the LCE. Gradually, the strain increases until it eventually reaches a steady state. After some time, an electrical short path is provided to the electrodes of the DLCEA by a custom switching mechanism. The DLCEA is thus discharged and abruptly contracts elastically and  then continues to further contract slowly again due to the viscoelasticity (Fig. 2, fig. S8). In the case of  ${\bf n} || {\bf F_g}$, there is no appreciable actuation along the loading direction due to the substantially higher stiffness (Movie S1) therefore no further tests were conducted on this configuration DLCEA.

Simultaneously with the actuation, a high speed camera (Vision Research v641) is hand triggered. For the contraction data presented in the main text (Fig 2B), the camera acquires at 1400~FPS. The video frame during which the high voltage is switched was identified by a pair of LEDs triggered by the same solid state relay as the high voltage switches. The discharge current is measured across a resistor divider pair in series with the DLCEA by an oscilloscope (Textronix MDO4024C). A schematic of the high voltage switching mechanism used to measure discharge current by reading the voltage, V$_\mathrm{D}$, across a known resistor, R$_\mathrm{D}$, is shown in Figure~S8A. The actuation also depends on the applied voltages and for each load, voltages of 2~kV, 2.5~kV, and 3~kV were tested (Fig.~S8C).

Video data from the contractile tests was tracked with Tracker Video Analysis \cite{TrackerVideoAnalysis}, and then analyzed in custom Python scripts. In these tests, the high voltage was switched on for approximately 20~s prior to discharge so that the DLCEA would reach its actuated stable rest length.  Initial distances were marked by hand in Tracker Analysis and then compared to known component sizes to compute distances and subsequently energy, power, and efficiency measurements. Oscilloscope data was also analyzed in custom Python code. A baseline capacitive charge was subtracted from the measured discharge by measuring the discharge with no DLCEA attached to the switch -- each meter of high voltage cable has a capacitance of approximately 100~pF.

\subsection*{Uniaxial buckling DLCEA, frequency response characterization}

A uniaxial LCE film is constrained at its edges by a laser cut Plexiglas frame. The film is carefully placed on top of the frame so as not to induce prestrain or leave any slack. A central square carbon grease electrode is painted onto the film on both sides through a low  adhesive removable tape mask. The in-plane length of the film grows along the soft direction, but, due to the fixed boundary conditions, it creates a buckled wrinkling pattern. The height of the wrinkle pattern was measured by laser confocal profilometry in the off-state and every 250~V starting from 500~V to 2.5~kV. In the 2.5~kV activated state, the out-of-plane peak-to-peak stroke is 1.47~mm or 1800\% of the LCE film thickness which is approximately 80 \textmu m.

To determined the frequency response of the uniaxial buckling DLCEA, we applied a sinusoidally varying 1~kV supplied by a Physik Instrument E-107 piezo high voltage amplifier. The input signal was generated by a function generator (Textronix ). The motion of the DLCEA membrane was observed by a Thorlabs Telesto optical coherence tomography microscope. The DLCEA film maximum height was first found manually in the DC on-state and subsequently observed in the same location for various frequencies (Fig 3D, fig. S9).

\subsection*{Programmed buckling DLCEA}

The defect array was achieved by programming light polarization from laser cut squares of a linear polarizer film which was stitched back into the desired grid on a glass slide using NOA~65 UV curing glue. The stitched pieces of polarizing squares are not perfectly beside each other due to imperfections in the laser cutting step and difficulty in manual stitching. However, the unaligned boundaries between aligned regions in the LCE film are small and do not apparently affect the actuation response.

Similar to the uniaxial buckling, after fabrication of the LCE and removing it from the BY coated glass slides, we fixed the LCE film to a laser cut Plexiglas frame and covered the programmed region with electrically conductive carbon grease. The programmed region is easily distinguished under ambient lighting from the surrounding areas due to the mismatch in index of refraction between isotropic and well aligned regions.

\subsection*{Twisted nematic DLCEA}

Twisting electric actuation in LCEs has been observed in smectic films due to the electroclinic effect \cite{SpillmannElectricallyInducedTwist2016}. In an approximately 90 degree twisted nematic configuration, the LCE director rotates through the thickness of the LCE film thus breaking front-back symmetry. When actuated, the perpendicular director orientation between the front and back faces creates perpendicular preferred expansion directions on the opposing faces. The exact activated form of a twisted nematic DLCEA is dependent on the geometry of the system \cite{SawaShapeselectiontwistnematicelastomer2011}. We tested two twisted nematic configurations by cutting the twisted nematic LCE film at different angles with respect to the director alignment in the mid-plane of the film. When the mid-plane director is parallel to the films' long axis, $\mathbf{n}_{\mathrm{mp}} || l$, the actuated DLCEA twists into a helical shape as in Figure~S10A. When $\mathbf{n}_{\mathrm{mp}} \perp l$, as in Figure~S10B the film begins to spiral around the mid-plane director but is restricted by gravity. The unconstrained active twist angle for the case $\mathbf{n}_{\mathrm{mp}} || l$ is much larger than $\mathbf{n}_{\mathrm{mp}} \perp l$ (Fig. S10C). We also clamp the free ends in a rheometer to measure the torques they produce when actuated (Fig. S10C). We anticipate stacked multilayer or larger aspect ratio DLCEAs may produce significantly larger torques.
 
\clearpage

\section*{Supplementary figures}

\begin{center}
\includegraphics[width=120mm]{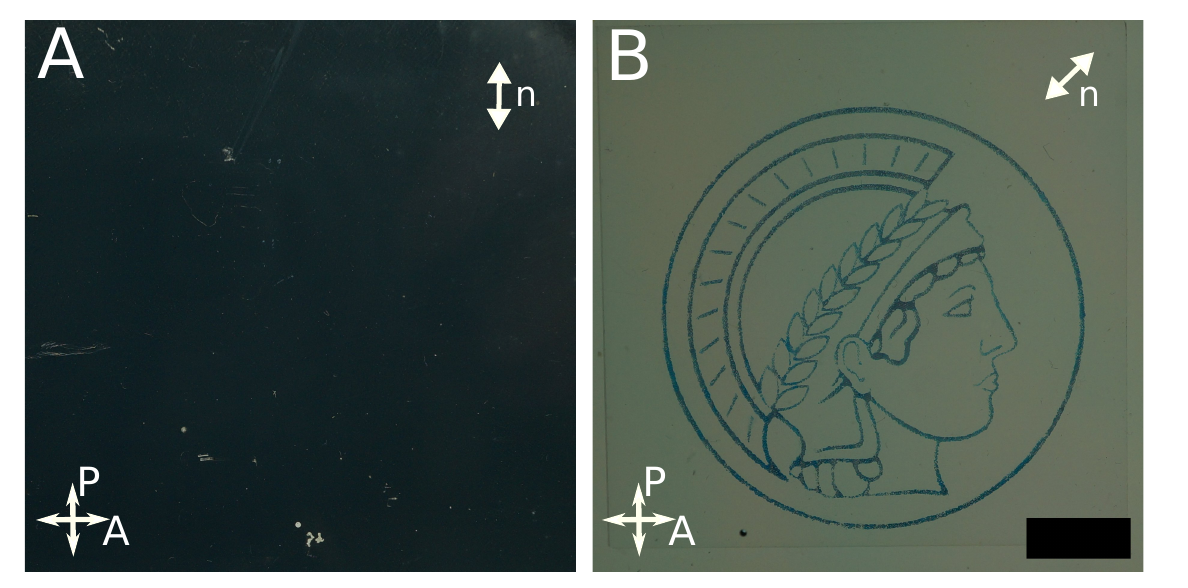}
\end{center}

\noindent {\bf Fig. S1. Optical characterization of uniaxial LCE.}
Large areas of defect free uniaxially aligned LCE. Polarized optical images demonstrate high uniformity large area fabrication of LCE with uniaxial alignment with {\bf n} parallel (\textbf{A}) to the polarizer ({\it P}) and {\bf n} (\textbf{B}) aligned 45$^{\circ}$ to the polarizer. The image of Minerva stamped on glass is present behind the polarizer in both images, and the high contrast between director orientations indicates excellent alignment. The scale bar is 1 centimeter.

\clearpage
\includegraphics[width=150mm]{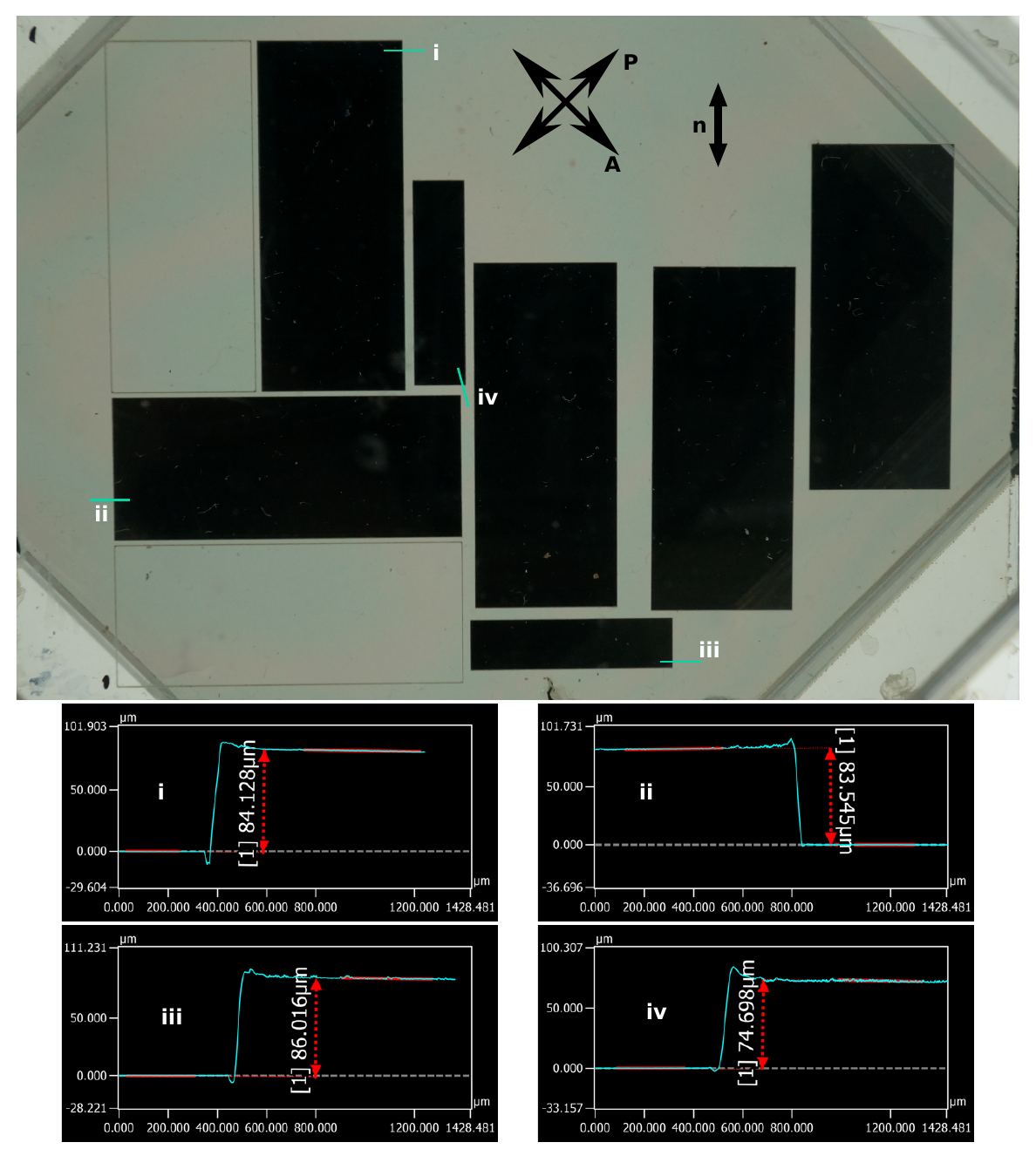}

\noindent {\bf Fig. S2. Photograph of laser cut regions of LCE and measured heights.} Representative LCE film with laser cut regions removed for DLCEA assembly. The film thickness measured at several points by laser confocal profilometry shows small variations in film thickness.

\clearpage

\begin{center}
\includegraphics[width=120mm]{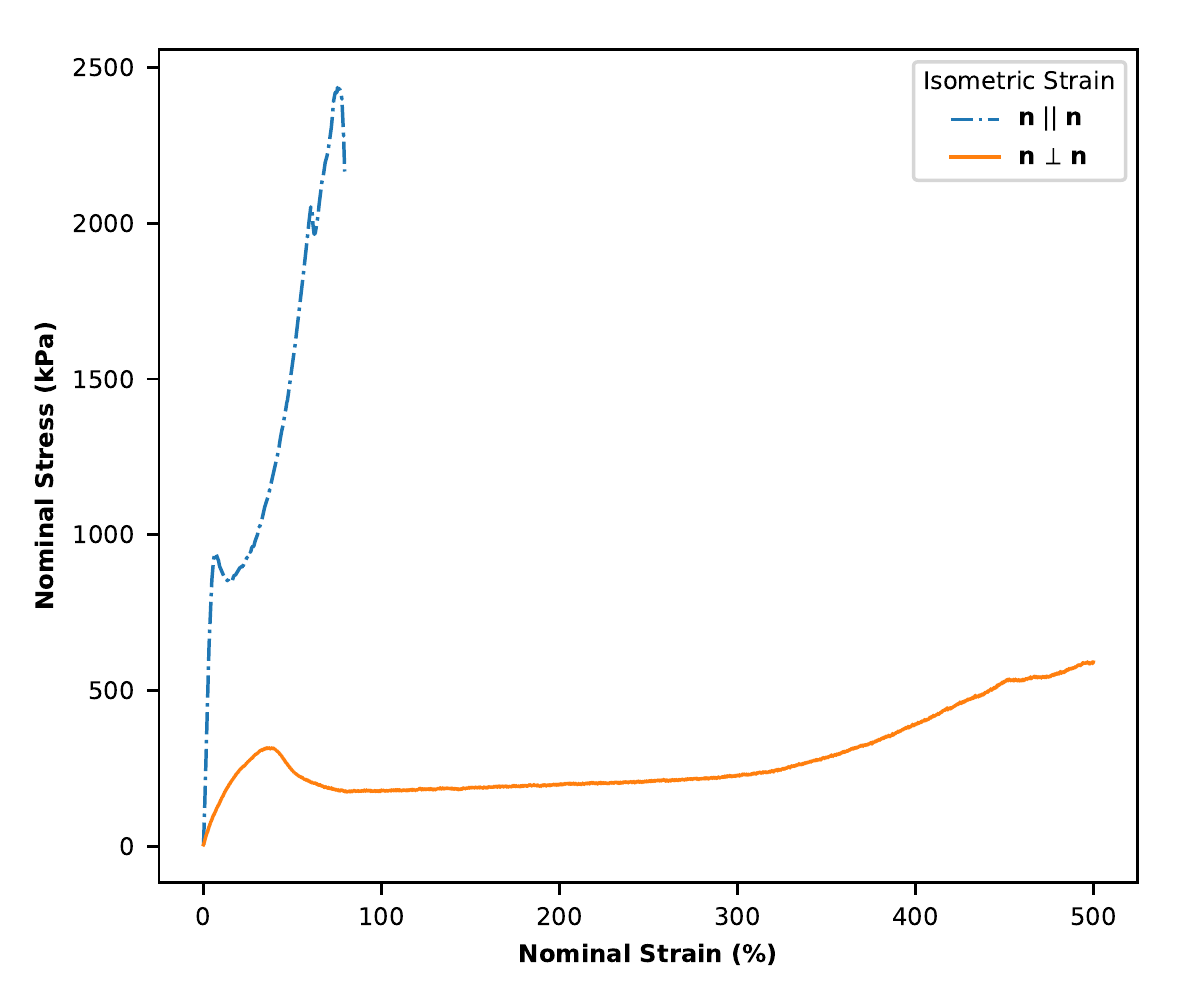}
\end{center}

\noindent {\bf Fig. S3. Stress-strain characterization of uniaxial LCE.}
Test pieces cut from neighboring regions of the samples used to make DLCEAs are tested for their high strain response. The sample with the director perpendicular to the strain (${\bf n} \perp {\bf u}$) exhibits a critical strain and the onset of director reorientation at approximately 50\% nominal strain.

\clearpage

\begin{center}
\includegraphics[width=150mm]{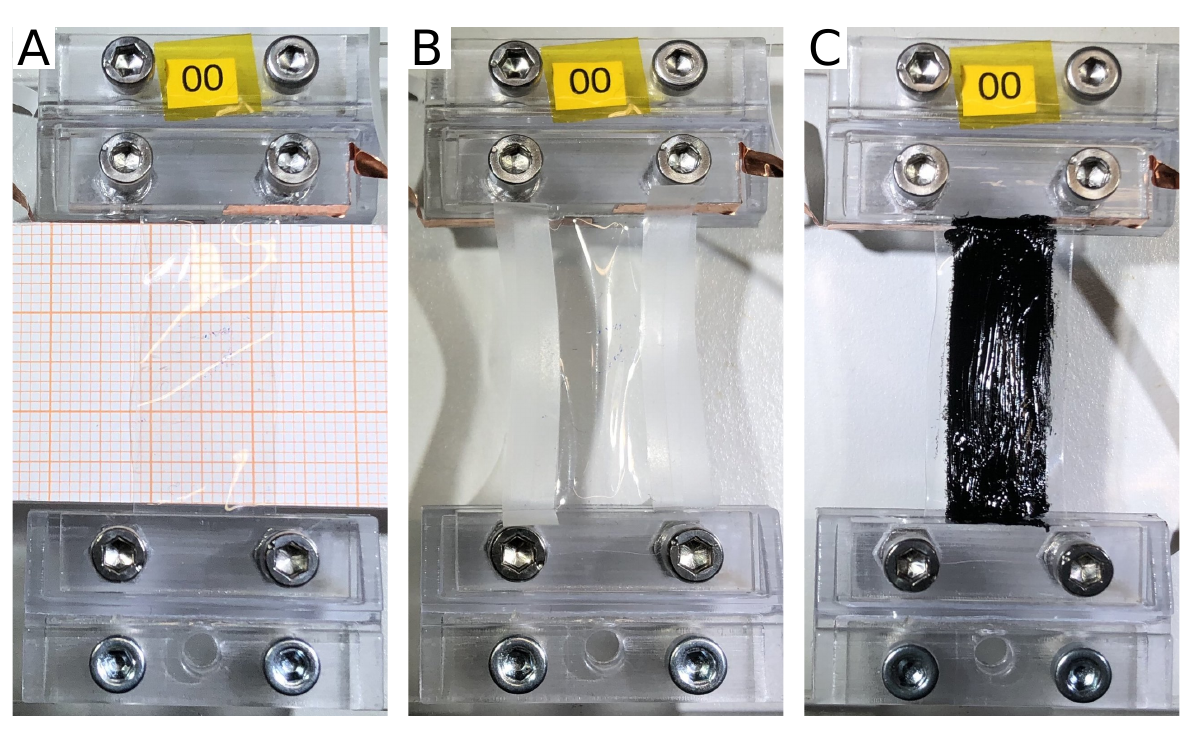}
\end{center}

\noindent {\bf Fig. S4. Assembly process for typical uniaxial DLCEA build.} ({\bf A}) The LCE film is aligned and clamped in a plastic holder in front of a millimetre paper scale. ({\bf B}) The edges of the LCE film are masked with tape to cover two millimeters on either edge. ({\bf C}) The carbon grease is painted on to the LCE and the tape mask is removed.

\clearpage

\begin{center}
\includegraphics[width=120mm]{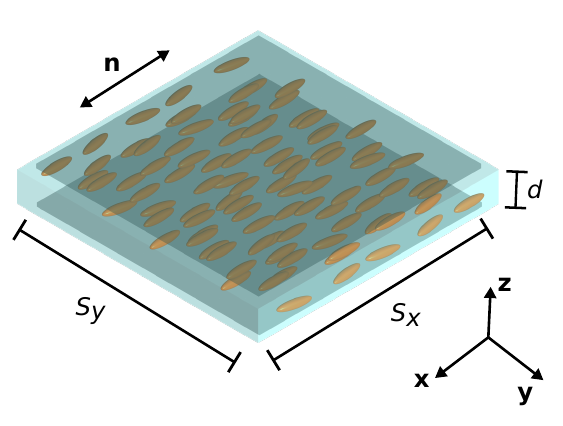}
\end{center}

\noindent {\bf Fig. S5. Schematic of DLCEA with coordinate axes.}
The director, ${\bf n}$, is parallel to the $\bf{x}$-axis and defines the stiff direction of the LCE film and the average direction of the rod-like LCE molecules. The $\bf{y}-\bf{z}$ plane is isotropic.

\clearpage

\begin{center}
\includegraphics[width=120mm]{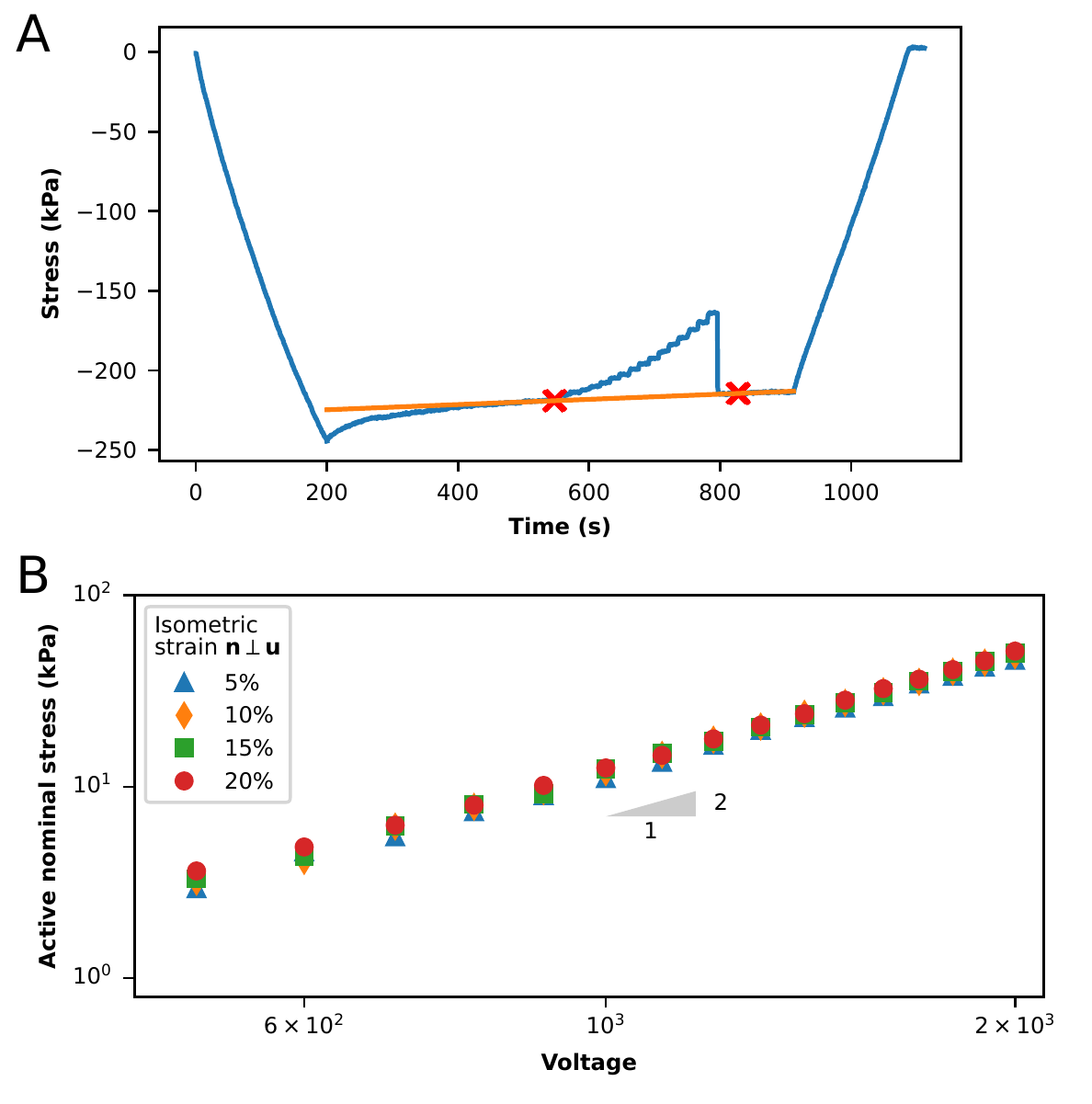}    
\end{center}

\noindent {\bf Fig. S6. Isometric uniaxial DLCEA relaxation and log-log stress voltage relation.} {\bf (A)} Active stresses in a DLCEA with 20\% isometric strain and ${\bf n} \perp {\bf u}$ were found after a relaxation period and by subtracting a linear fit baseline. {\bf (B)} Log-log plot of active stress in an isometric configuration reveals the quadratic relationship between actuation voltage and Maxwell stress.

\clearpage

\begin{center}
\includegraphics[width=120mm]{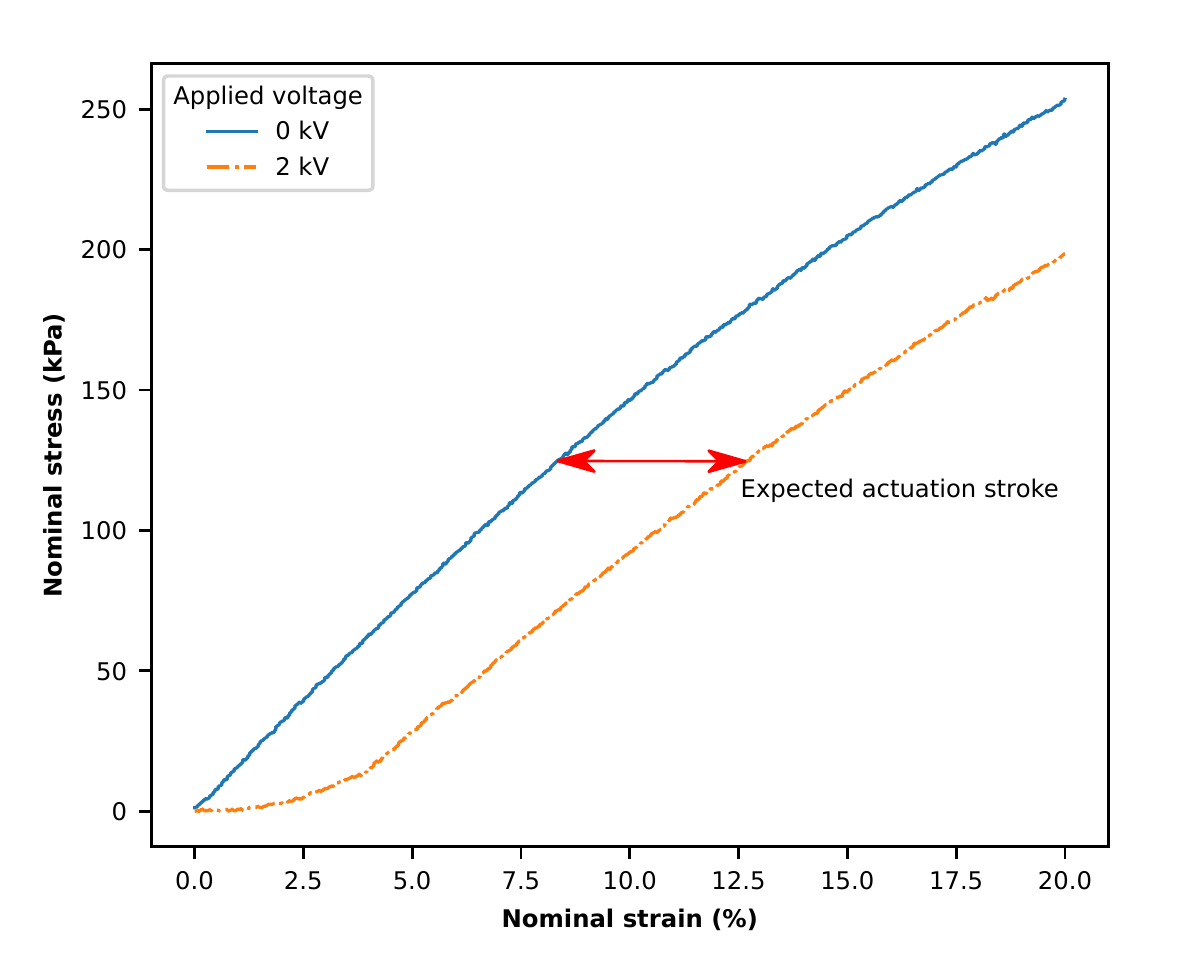}
\end{center}

\noindent {\bf Fig. S7. Isopotential tests of uniaxial DLCEA.}
The shift in the stress-strain curve at 0~kV and 2~kV actuation voltage indicates the expected stroke when the DLCEA is actuated with a constant load.

\clearpage

\begin{center}
\includegraphics[width=150mm]{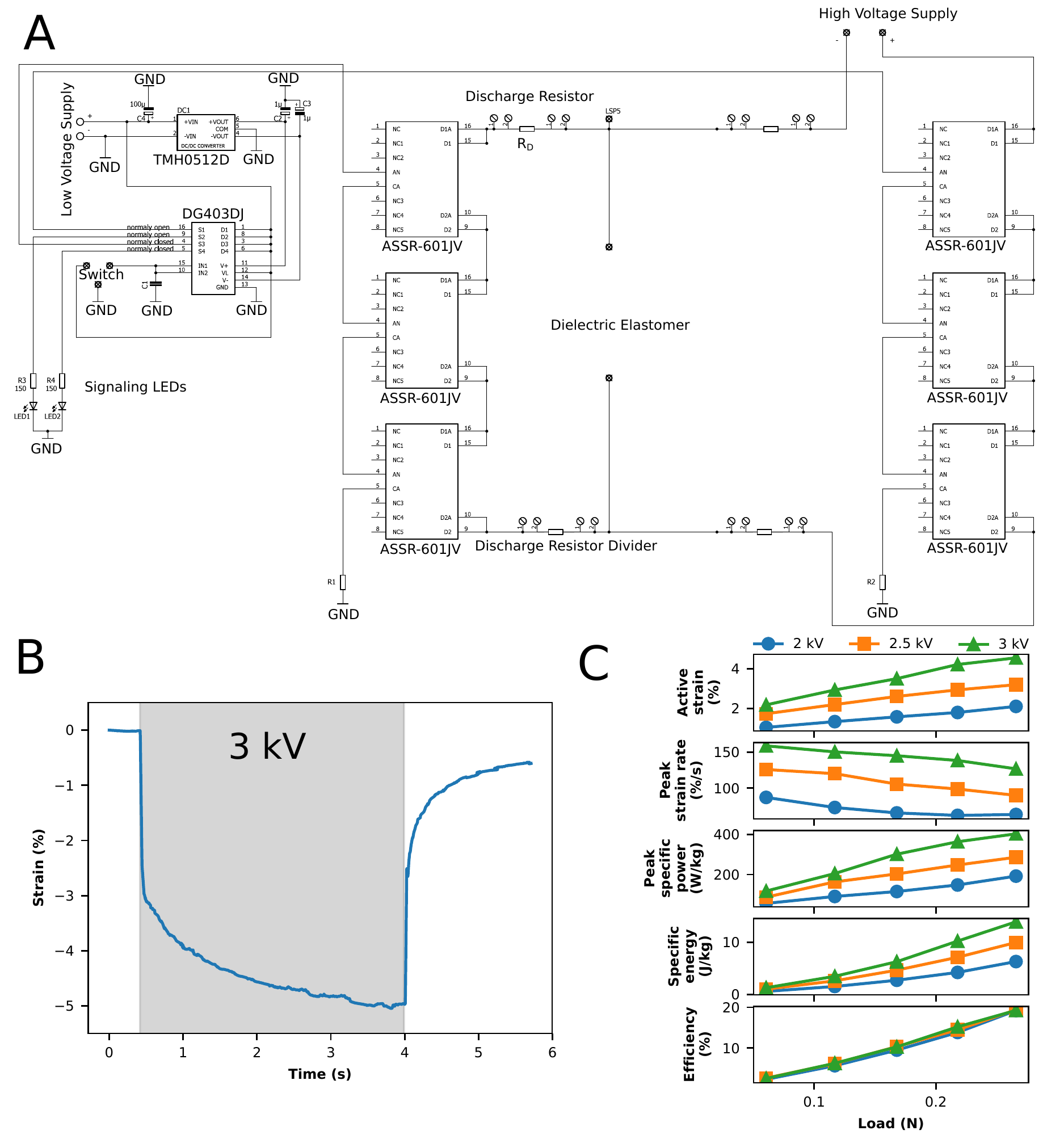}
\end{center}

\noindent {\bf Fig. S8. Schematic of high voltage switching mechanism,  isotonic full cycle actuation, and isotonic actuation characteristics with varying voltage.} {\bf (A)} The custom high voltage switch uses a two channel low voltage solid state relay (DG403DJ) to control both the high voltage relays (ASSR-601JV) and the signaling LEDs used in high speed video. {\bf (B)} An example fully cycle actuation of a DLCEA showing significant creep following an elastic response. {\bf (C)} The istonic actuation results vary with the applied actuation voltage.

\clearpage

\begin{center}
\includegraphics[width=150mm]{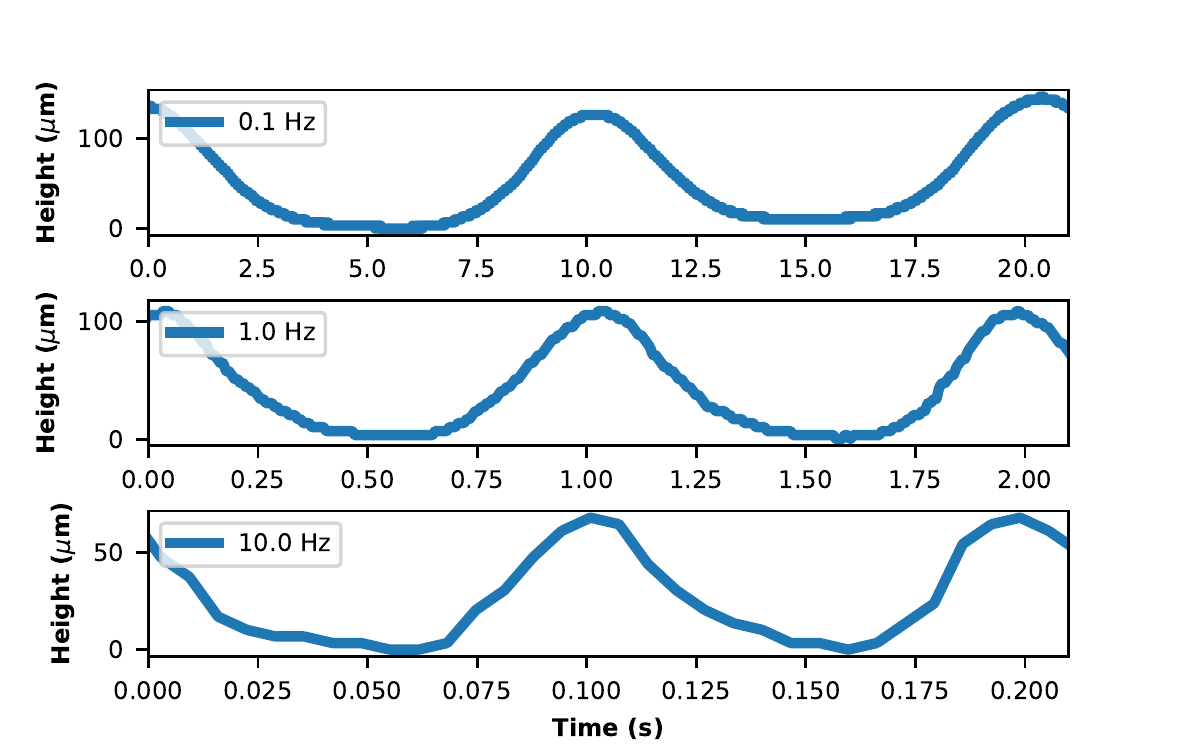}
\end{center}

\noindent {\bf Fig. S9. Uniaxial buckling DLCEA frequency response.} The time series height near an anti-node on a buckling uniaxial DLCEA driven at 1~kV for 0.1~Hz, 1.0~Hz, and 10.0~Hz driving frequencies.

\clearpage

\begin{center}
\includegraphics[width=120mm]{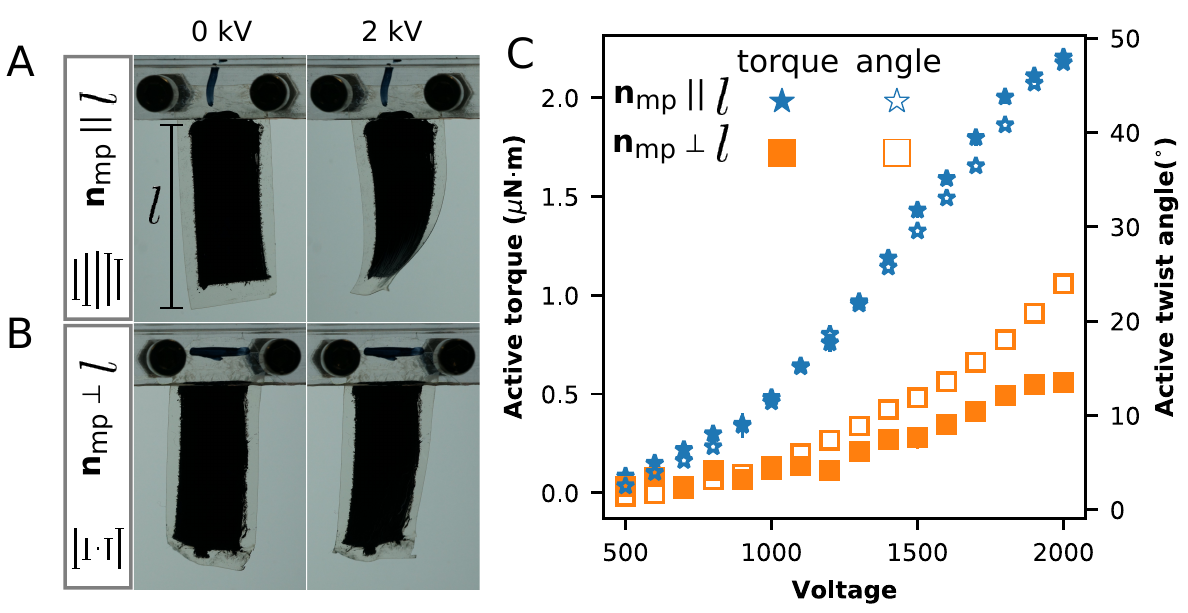}
\end{center}

\noindent {\bf Fig. S10. Twisted nematic DLCEA.} Actuation of LCE films with twisted nematic configuration. The director twists $90^{\circ}$ through the LCE film thickness. (\textbf{A}) Off and 2 kV actuated states for the mid-plane director parallel to the long axis of the film, $\mathbf{n}_{\mathrm{mp}} || l$. (\textbf{B}) Off and 2 kV actuated states for the mid-plane director perpendicular to the long axis of the film, $\mathbf{n}_{\mathrm{mp}} \perp l$.  (\textbf{C}) Quasistatic high voltage induced twisting and torques. Active torque was measured by a rheometer with torsional clamps. Active twist was measured using image tracking of the films' free end.

\clearpage

\subsection*{Other supplementary material for this manuscript includes the following:}
Movies S1 to S4

\section*{Supplementary movie captions}

\subsection*{Movie S1}
\noindent {\bf Uniaxial dielectric liquid crystal elastomer actuator with director parallel to  $\bf{F_g}$.} On a uniaxial DLCEA with ${\bf n} || {\bf F_g}$ and a load of 0.22~N, a 3~kV potential and discharge path is cycled alternately.

\subsection*{Movie S2}
\noindent {\bf Uniaxial dielectric liquid crystal elastomer actuator with director perpendicular to $\bf{F_g}$.} A uniaxial DLCEA with ${\bf n} \perp {\bf F_g}$ and a load of 0.06~N, a 3~kV potential is applied to charge the electrodes and then switched to provide the electrodes a discharge path.

\subsection*{Movie S3}
\noindent {\bf Demonstration of uniaxial buckling dielectric liquid crystal elastomer actuator.} A uniaxially aligned DLCEA buckles when charged by a 2.5~kV potential and then flattens when the electrodes are discharged. The string held taught over the DLCEA helps to visualize the actuation but is also slightly displaced by the buckling.

\subsection*{Movie S4}
\noindent {\bf Demonstration of programmable shape change buckling dielectric liquid crystal elastomer actuator.} A DLCEA, locally aligned by photo programming, buckles into a programmed shape configuration when charged by a 2.5~kV potential and then flattens when the electrodes are discharged.

\clearpage



\clearpage

\end{document}